# MapExif: an image scanning and mapping tool for investigators

**Lionel Prat**
**NhienAn LeKhac**
**Cheryl Baker**
*University College Dublin, Belfield, Dublin 4, Ireland*

Abstract

Recently, the integration of geographical coordinates into a picture has become more and more popular. Indeed almost all smartphones and many cameras today have a built-in GPS receiver that stores the location information in the Exif header when a picture is taken. Although the automatic embedding of geotags in pictures is often ignored by smart phone users as it can lead to endless discussions about privacy implications, these geotags could be really useful for investigators in analysing criminal activity. Currently, there are many free tools as well as commercial tools available in the market that can help computer forensics investigators to cover a wide range of geographic information related to criminal scenes or activities. However, there are not specific forensic tools available to deal with the geolocation of pictures taken by smart phones or cameras. In this paper, we propose and develop an image scanning and mapping tool for investigators. This tool scans all the files in a given directory and then displays particular photos based on optional filters (date/time/device/localisation…) on Google Map. The file scanning process is not based on the file extension but its header. This tool can also show efficiently to users if there is more than one image on the map with the same GPS coordinates, or even if there are images with no GPS coordinates taken by the same device in the same timeline. Moreover, this new tool is portable; investigators can run it on any operating system without any installation. Another useful feature is to be able to work in a read-only environment, so that forensic results will not be modified. We also present and evaluate this tool real world application in this paper.

*Keywords:*

MapExif: an image scanning and mapping tool for investigators

**Introduction**

Since the start of the digital information age to the rise of the Internet, the amount of digital data has dramatically increased. Some data is structured and stored in a traditional relational database, while other data, including pictures and videos, is unstructured. Organizations also have to consider new sources of data generated by new devices such as smart phones and cameras. The availability and adoption of newer, more powerful mobile devices, coupled with ubiquitous access to global networks will drive the creation of more new sources for data. As a consequence, we have had a deluge of data from not only science and industry fields but also digital forensics fields. Obviously, handling and analysing large and complex data has always offered the greatest challenges as well as benefits for organisations of all sizes. Although the amount of data available to us is constantly increasing, our ability to process it becomes more and more difficult. This is especially true for the criminal investigation today.

One of the big challenges for the IT forensics department of the French Gendarmerie Forensic Sciences Institute is the number of pictures investigators have to analyse. Sometimes it is not urgent as for instance, the suspect is already in prison, and a traditional image forensics approach can be applied. However, most of the time, especially for field investigators, the images retrieved need to be analysed in a very short time. The following example shows the requirement for a geolocation-based tool. A floating body is discovered in a river and a "friend" of the victim is suspected. During searching, the investigators found a number of smart phones, hard disks, memory cards… The suspect claims to have not been at the crime scene. To verify his story, investigators have to analyse thousands of photos by manually checking the geolocation data embedded into the picture. During the analysis they discovered a picture of him in an area, near the crime scene. Unfortunately it was 48 hours before the picture was found, and due to lack of proof, the suspect had already been released. This example demonstrates the need for a tool that helps the mapping and analysing of photos based on exif data, especially geotagging.

In fact, there are software available nowadays, such as GeoSetter (GeoSetter, 2014) andGoogle Picasa (Google Picasa, 2014), that can display the position of pictures on a map. However, they just spot all images regardless of the device used, and with no filter function. The only way to have a custom view for the user is to select which images to display before they are shown on the map. Moreover, Geosetter and Google Picasaare not forensics tools; investigators do not know if these tools make changes to the data, which would be unacceptable if they were to be used as evidence. These current tools moreover cannot handle images without GPS information; neither to generate user reports.

Therefore, in this paper, we present a new geolocation-based forensics tool, MapExif, that scans all pictures retrieved from smart phones, cameras, etc. and extracts the exif data (date/time, device, geolocation) to display them efficiently on a map. MapExif is developed in PHP/MySQL, thereby making it compatible with different operating systems. It can run on both a single computer and on a server. It scans recursively all files in a given directory and identifies and indexes all the images in the databases. MapExif also offers important functionalities such as zone filter, device filter, and date filter. Indeed, this tool can handle and display images without GPS information. It makes the connection between a geotagged non-geotagged images taken by the same device in the same time line. As pictures do not always have geotagged information, for example in the case of a photograh taken inside a building, this functionality is very useful. By using MapExif, investigators can also generate reports and easily manipulate the results without fear that the results could be altered. We describe in this paper the architecture and the development of this tool. We also discuss experiments using MapExif tool in scanning and mapping forensics images during real work application.

The rest of the paper is organised as follows: Section 2 describes the background research in the area. Section 3 details the architecture of the MapExif tool. We present and analyse its functionalities in Section 4. Next, we discuss the test results of the tool in different forensics scenarios in Section 5. Finally, we conclude in Section 6.

**Background**

In this section, we present firstly the background concepts related to photographic metadata, exif format and geotagging. We then look at related work in this context.

**Photographic Meta-data**

The term metadata refers to "data about data". It provides information about certain item's content. Metadata may be written into a digital photo file that will identify who owns it, copyright & contact information, what camera created the file, along with exposure information and descriptive information such as keywords about the photo, and of most relevance to this paper the geolocation information, making the file searchable on the computer and/or the Internet. The camera can write Metadata to the pictures at the time of taking the photograph, or sometimes it is done by the photographer and/or software after downloading the image to a computer.

Photographic Metadata Standards have been developed by a number of organizations. They include, but are not limited to:

- IPTC Information Interchange Model IIM (International Press Telecommunications Council)

- IPTC Core Schema for XMP

- XMP – Extensible Metadata Platform (an ISO standard)

- Exif – Exchangeable image file format, Maintained by CIPA (Camera & Imaging Products Association) and published by JEITA (Japan Electronics and Information Technology Industries Association)

- Dublin Core (Dublin Core Metadata Initiative – DCMI)

- PLUS (Picture Licensing Universal System).

**Exif**

EXIF (Exif, 2010) stands for "Exchangeable Image File Format". This type of information is formatted according to the TIFF specification, and may be found in JPG, TIFF, PNG, JP2, PGF, MIFF, HDP, PSP and XCF images, as well as many TIFF-based RAW images, and even some AVI and MOV videos.

The first JPEG file format standard (JFIF) was defined in 1992 to enable the interchange of JPEG bit streams between a wide variety of applications and platforms. This was improved in 1998 to a new standard EXIF (Exchangeable Image File Format), allowing camera manufacturers to embed and store camera and image metadata into JPEG and TIFF files. This metadata can give the forensic investigator the ability to extract vital evidences, such as when the picture was taken, what camera was used in capturing the image, who took the image and where the image was captured. EXIF metadata standard also enables digital camera manufacturers to include directly on the image file, information such as camera make and model, camera settings, time, author and  copyright Therefore the photographer has a permanent record of this

MAPEXIF: AN IMAGE SCANNING AND MAPPING TOOL                                                   7information preserved along with the image. For instance, JPEG file begins with "FFD8" which is the SOI (Start of Image) Marker and ends with "FFD9" which is the EOI (End of Image) marker. Between these two markers, the data is divided into different segments, defined by a specific marker for each segment. The flexibility of the file structure gives the ability to add more segments and markers while still conforming to the JPEG specification. The following link lists all EXIF tags: http://www.sno.phy.queensu.ca/~phil/exiftool/TagNames/EXIF.html

 Also listed are TIFF, DNG, HDP and other tags that are not part of the EXIF specification, but may co-exist with EXIF tags in some images. Tags, which are part of the EXIF 2.3 specification, have an underlined Tag Name.

**Geotagging**

Geotagging is the process of adding geographical identification metadata to various media such as a photograph or video, websites, SMS messages, QR Codes or RSS feeds, and is a form of geospatial metadata. This data usually consists of latitude and longitude coordinates, although it can also include altitude, bearing, distance, accuracy data, and place names.

Geo-tagging uses, by descending order of accuracy, GPS, WiFi router locations, or cell tower identification to spot the position on the globe. A geotagged photograph is a photograph that is associated with a geographical location by geotagging. Usually this is done by assigning at minimum, a latitude and a longitude to the image. Altitude, compass bearing and other fields may also be included.

There are two main options for geotagging photos: capturing GPS information at the time the photo is taken, or "attaching" the photograph to a map after the picture is taken. In order to capture GPS data at the time the photograph is captured, the user must either have a camera with built in GPS, or a standalone GPS along with a digital camera. Because of the requirement since September 11, 2012 for wireless service providers in United States to supply precise location information for 911 calls by, more and more cell phones have built-in GPS chips. Most smart phones already use a GPS chip along with built-in cameras to allow users to automatically geotag photos. A few digital cameras also have built-on or built-in GPS that allow for automatic geotagging. Devices use GPS, A-GPS or both. A-GPS may get a faster initial fix if you are within range of a cell phone tower, and may work better inside buildings.

**Geotagging in JPEG exif metadata**

As seen previously, the Exif format has standard tags for location information. Most of the smartphones and some cameras have a built-in GPS receiver that stores the location information in the Exif header when a picture is taken. Some other cameras have a separate GPS receiver that fits into the flash connector or hot shoe. Recorded GPS data can also be added to any digital photograph on a computer, either by correlating the time stamps of the photographs with a GPS record from a hand-held GPS receiver, or manually by using a map or mapping software. The process of adding geographic information to a photograph is known as geotagging. Photo-sharing communities like Panoramio, locr or Flickr equally allow their users to upload geocoded pictures or to add geolocation information online. Exif data is embedded within the image file itself. While many recent image manipulation programs recognize and preserve Exif data when writing to a modified image, it is not possible to completely rely on the metadata as it can be easily modified.

With photos stored in JPEG file format, the geotag information is typically embedded in the metadata (stored in Exif or XMP (Extensible Metadata Platform) format). Generally Exif metadata is technical information about the image captured by the digital camera while XMP metadata is information specific to the photographer himself. These data are not visible in the picture itself but are read and written by special programs such as the cross-platform open source ExifTool11, most digital cameras and modern scanners.

When stored in Exif, the coordinates are represented as series of rational numbers in the GPS sub-IFD (Fig.1). Latitude and longitude are stored in units of degrees with decimals (Fig.2) or the same coordinates could also be presented as decimal degrees (the necessary format to fit with Google Maps API, Fig.3). This geo-tag information can be read by many programs, such as the cross-platform open source ExifTool.

```
+ [GPS directory with 5 entries]
| 0) GPSVersionID = 2 2 0 0
| - Tag 0x0000 (4 bytes, int8u[4]):
| dump: 02 02 00 00
| 1) GPSLatitudeRef = N
| - Tag 0x0001 (2 bytes, string[2]):
| dump: 4e 00 [ASCII "N\0"]
| 2) GPSLatitude = 57 38 56.83 (57/1 38/1 5683/100)
| - Tag 0x0002 (24 bytes, rational64u[3]):
| dump: 00 00 00 39 00 00 00 01 00 00 00 26 00 00 00 01
| dump: 00 00 16 33 00 00 00 64
| 3) GPSLongitudeRef = W
| - Tag 0x0003 (2 bytes, string[2]):
| dump: 57 00 [ASCII "W\0"]
| 4) GPSLongitude = 10 24 26.79 (10/1 24/1 2679/100)
| - Tag 0x0004 (24 bytes, rational64u[3]):
| dump: 00 00 00 0a 00 00 00 01 00 00 00 18 00 00 00 01
| dump: 00 00 0a 77 00 00 00 64
```

*Figure 1*. Hexadecimal dump of the relevant section of the Exif metadata (with big-endian order)

```
GPS Latitude  : 57 deg 38' 56.83" N
GPS Longitude : 10 deg 24' 26.79" E
GPS Position  : 57 deg 38' 56.83" N, 10 deg 24' 26.79" E
```

*Figure 2*. An example of GPS information readout for a photo

```
GPS Latitude  : 57.64911
GPS Longitude : 10.40744
GPS Position  : 57.64911 10.40744
```

*Figure 3*. GPS information presented as decimal degrees

.

**Geotagging in JPEG exif metadata**

While GPS data in Exif metadata is becoming more frequent, without the user being always aware (Friedland et al., 2010): the question arises whether we can rely on it as part of an investigation.

There are many Google Tools that can assist investigations, such as Google Maps, Google Picasa, Google Realtime andGoogle Reader. These tools are normally the investigator's first choice in searching or mapping relevant information and/or pictures. Among these tools, Google Maps and Google Picasa can be used to deal with geo-location and geo-tagging of pictures taken by smart phones or camera. Google Maps, for instance, allows you to plot any number of locations and how these look from a satellite. Indeed, the Street View tool offers valuable location intelligence, as it can provide a virtually 360-degree view of any location including buildings and landscaping. For example the NYPD used Street View images in a prosecution of a drug case (Olivarez-Giles, 2010). However, despite the fact that some investigators do use Street View when conducting forensic investigations, and may even use it as evidence in court-, it is not a forensics tool and it does not have any imaging service that would enable the sorting of pictures of suspicious characters, geotagging, etc.

Google Picasa is an image-sharing service, which allows users to upload pictures, create albums and add geotagging. Investigators can use this tool to search for "tags" or labels including names or descriptions andthey can also determine the image's location by viewing its geotags. However Picasa just spot all images regardless of the device used and with no possibilities of filters on the map.

GeoSetter (GeoSetter, 2014) is a program that allows users to add/edit metadata to digital images. It allows you to "tag" images with location and other keywords for better and easier image searching.Having the ability to tag photos makes the way we store and search for images more user-friendly. GeoSetter, however is not a forensics tool; investigators do not know if it makes changes to the data, which would be unacceptable for evidence.

GeoJot (GeoJot, 2014) is a tool that helps the user to organise his photos. One of the main features is the ability to edit picture information and modify their geo-positions on a map. It also has the ability to merge information from a GPS track log (GPX file) and a non-geotagged picture in order to put it on the map. It is the opposite of a forensic tool, as the main function is to edit pictures.

There have been studies on the vulnerabilities of geo-tagging technology., The study "Challenging the Reliability of iPhone Geo-tags" (Lallie, 2011) shows two methods where the geographic coordinates of a picture taken with an iPhone can be modified therefore demonstrating that the information is unreliable. Furthermore the accuracy and the reliability of evidence extracted from GPS memory systems can be questioned for three main reasons:

- Dilution of precision (DOP) of GPS data
- GPS jamming (Last, 2009, Grant 2009)
- Improper legal engineering

GPS receivers can also be "spoof-attacked" wherein a signal is generated making the GPS receiver "believe that it is in motion" (Iqbal, 2008) and in return results in unreliable coordinates that may be recorded within the Exif data. As is has been demonstrated that the devices and the data contained within can be tampered with, this could lead to claims that the data storage systems cannot be proved to be secure (Iqbal, 2008).

Another study, "Metadata Based Forensic Analysis of Digital Information in the Web" (Salama, 2012), highlighted that metadata in general can be easily modified, and emphasized the need to protect metadata if it is to be used to provide reliable information for forensic investigations.

Regarding vulnerabilities, some of them are explored by C. Strawn (Strawn, 2009) and are remedied in part by technologies such as Assisted GPS (AGPS) (Djuknic, 2001, Feng, 2002, Zandbergen, 2009).

## MapExif Architecture

In this section, we describe the architecture of MapExif (Fig. 4). In fact, MapExif is composed of three main components: database management, image scanning/indexing and visualisation.

**Image scanning and indexing**

Starting from pictures, the idea is to extract metadata that is useful for the investigator (geolocation, timestamp, serial-number…), pre-process this information and then populate a database. The goal is to select the relevant images to show to the investigators. Images displayed on the map will be the ones which match with the database query. So the data embedded into the picture is very important, it is collected, processed and then displayed.

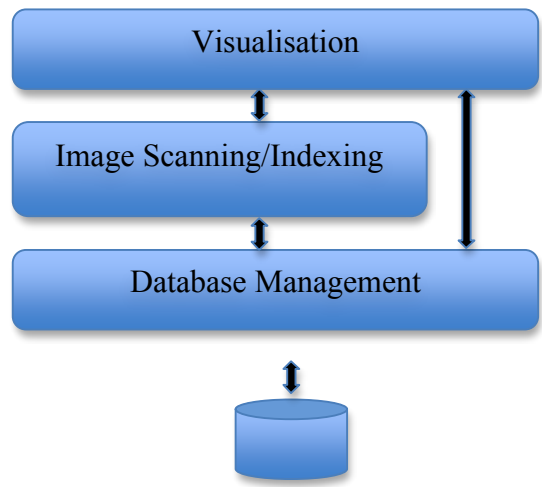

*Figure 4.* MapExif Architecture

The scanning and indexing of images consists of different stages: First of all, all image files are classified into two groups: those with and without geo-tagged information. Secondly, geo-location data of geo-tagged images is extracted. Images with the same coordinates are tagged (as only one spot will be displayed on the map, the investigator needs to know whether there are other pictures at the same place), and reverse geocoding (obtaining the postal address based on geo-location data) is applied to pictures. The postal address offers the ability for more search options at a later stage. The last operation is to search for links between geo-tagged and non-geo-tagged pictures. In fact this operation can be done later, but it's more efficient to pre-process this data during the scan operation so the displaying of the images will be faster. Finally, indexed images and their relevant information will be stored in databases.

Besides, another important task, which should be performed during the scanning and indexing of images, is to check the accuracy of the information relating to an image. The EXIF tag "GPSProcessingMethod" of an image could inform about the geo-positioning method used. The values of "GPSProcessingMethod" are "GPS", "CELLID", "WLAN" or "MANUAL". When "GPSProcessingMethod" value is GPS, the EXIF tag "GPSDOP" shows the accuracy of the measure (the lower the better). MapExif can also do cross verification. For example it can check the difference between the EXIF date of the image (which can be very easily modified) and the "GPSDateStamp" which is not commonly known and - if not modified – very precise (UTC time given by satellite). In case of a large difference (more than 24 hours, a warning is shown. MapExif can also correlate the geo-location position and the altitude. When a picture is geotagged, there is a strong chance that the information about the altitude is embedded also. MapExif checks whether the altitude complies with the altitude on this position (querying a map service).

**Database Management**

Indexed images will be stored in a relational database that can be implemented with different DBMSs and exploited by web services (PHP/AJAX). In addition, markers are automatically generated that facilite the process of visualisation.

**Database structure**

The MapExif database is composed of three tables: (i) analysis table, (ii) marker table and (iii) information table. The first table stores the information about the analysis itself (start/end time, number of files scanned, number of images found, and among them how many are geotagged). This table is populated with relevant data during the scanning step. If the end time is not set, that means the scanning process did not finish properly. The marker table stores all markers and all the linked information (position, metadata, reverse geocoded address…). During the scan some information is pre-computed: for example the number of relations between geotagged and non-geotagged images in a given time slot. This could be calculated later on the fly during the generation of markers on the map but it would take time. So to make the application reactive; time-consuming calculations, which would never change for a given analysis, are stored in the database. Finally, the information table stores information about the "identifiers" that are generated for each device. It is useful to display the different numbers and colors of the markers on the map. This information can be retrieved directly from the table "markers", but for performance reasons the use of a dedicated table containing only the "identifiers" is more efficient.

**Generating markers dynamically based on SQL queries**

Our database supports different filters from users when they access image information. We set up a process to generate markers automatically based on SQL queries. As shown in the following figure (Fig.5), our process consists of three steps: (i) Creating of the SQL query based on the filters; (ii) Sending the SQL statement to a PHP page which dynamically generates an XML page using the AJAX technology and (iii) Parsing the newly generated XML page to display the markers based on the Google Maps API.

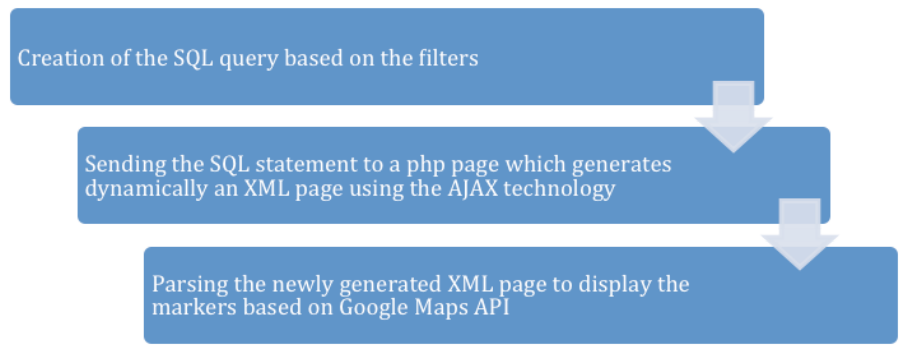

*Figure 5.* Generating makers

Indeed, Fig. 6 shows an example of the marker's table. This table includes the information related to a marker such as the unique id of each marker, name of the file, name of the generated thumbnail, timestamp added by the device when the picture was taken, timestamp caught with gps data which is useful to check if the datetime of the picture has been modified, generated fake id for the device to make it as unique as possible, order of the marker (number shown in the icon on the map), based on the fake_id, in descending order, number of non-geotagged pictures linked to the marker (1 stands for +/- 1 hour time slot) and all metadata which has been harvested.

Now that you have the table and the SQL statement, you can put them together to output the search results into an XML format that the map can retrieve through asynchronous JavaScript calls. Figure 7 is an example of XML output with PHP. The code for generating markers can be found in Appendix A.

*Figure 6* - Example of the marker's table

*Figure 7* - The outputted XML with PHP

**Visualisation**

MapExif is first of all a visualisation tool. The way the information is displayed on the map is very important. Therefore, MapExif is designed to support different levels of visualisation.

The first level of visualisation is to display images based on the basic filters of users such as device types, date, etc. The investigators will find the relevant images for their cases by setting up filters directly on the map.

The second level of visualisation allows users to define the area they want to explore in detail. MapExif can combine with underlying map application to display at street view level that allows investigator to explore more on the area where this picture taken.

MapExif supports another level of visualisation where it can combine non-geotagged images with geo- tag images based on certain conditions, such as whether they were taken in the same period. The aids (links with non-geotagged pictures…) are done automatically.

MapExif is based on a map application that allows the investigator to explore the images. One of the most important features of MapExif is portability. Therefore, MapExif can use both Google Maps and open-source software such as Open Street Map to implement its visualisation functionalities.. Investigators are not IT specialists and are not intended to be. Mapexif must find the most effective and most intuitive method to help.  The challenge is to not include too much automation, as to over automate can distort reality and lead to ambiguous conclusions.

More details on the visualisation of MapExif can be found in Section 4 and Section 5.

## MapExif Implementation

**Portable (Running MapExif on different OS)**

As most personal computers are running on Windows OS, we had to find a way to run a LAMP server (Linux/Apache/MySQL/PHP), which could work in a read-only environment such as a CD-ROM, without modifying anything on the machine. The solution found was to use the Server2Go16 platform. Thus with the autorun function, the user just has to insert the CD/DVD/USB … into the computer and the Apache/PHP/ MySQL servers automatically run in a pre-configured environment. It is also configured to automatically launch a portable version of Firefox, so there is no fear of a cross browsing issue. The Server2Go application automatically monitors the portable Firefox thread and automatically shuts down the servers when the browser is closed.

To run MapExif in a Linux environment, the user has just to put the PHP files into the Apache's default web root directory and the application automatically sets up. This procedure is not really user-friendly; another solution is the creation of a live CD based on Ubuntu with MapExif pre-installed.

**Image scanning and indexing**

As mentioned above, all files are scanned, and any JPG images are identified based on their header (Hex FFD8). In order to verify for each file if it is a JPEG image, MapExif uses the PHP function exif_imagetype()18 which reads the first bytes of an image and checks its signature. To extract metadata, MapExif uses the PHP function exif_read_data() which reads the EXIF headers from a JPEG or TIFF image file (and thus the metadata generated by digital cameras). Indeed, in this step, MapExif should identify the specific device used to take pictures. The idea is to extract all exif information relating to a specific device and create a "fake id" with the maximum of information. Most devices only record their brand and model in the exif metadata. If more information about the device is available, it is added to the fake id, so it is possible to distinguish between different devices. The codes for extracting GPS data as well as for identifying devices can be found in Appendix A.

Moreover, in this step, some pre-processing operations are required for the generation of markers in the following step. In order to display the markers efficiently, we perform some preprocessing tasks during the scan of the images such as:

- Sort by number of pictures per device in descending order
- Search for other pictures in the database having the same GPS coordinates
- Searchfor there are other pictures with the same time slot (plus or minus 1, 2, 3, 4, 5, 12, 24 hours)

As all this information is already stored in the database, it doesn't need to be calculated again, and therefore the process of exploiting the results is much faster. Figure 5 shows the interface of our scanning/indexing phase. It allows us to monitor the progress of this process.

**Database Design and Implementation**

The technique adopted is an interaction between PHP and a MySQL database. The markers are displayed with the Ajax technique so there is no need to reload the page when a filter (and then the corresponding SQL query) is modified. Indeed, PHP since version 5 integrates PDO (PHP Data Objects) that provides a data-access abstraction layer, regardless of which database is being used. For now, a MySQL database is used, but if the application will at a later stage be used more extensively, another database system can be chosen to handle big data. Figure 6 shows the structure of SQL database of MapExif.

**Visualisation**

As mentioned in Section 3, visualisation is the most important features of MapExif. In this subsection, we present the implementation of different functionalities related to the visualization. We start with the interface of MapExif. MapExif has two main functions: scanning and exploiting the results. When sent to a magistrate for instance, only the exploitation of the results is available. Figure 7 shows the main interface of MapExif that consists of four groups: filtering, status, live report and display zone.

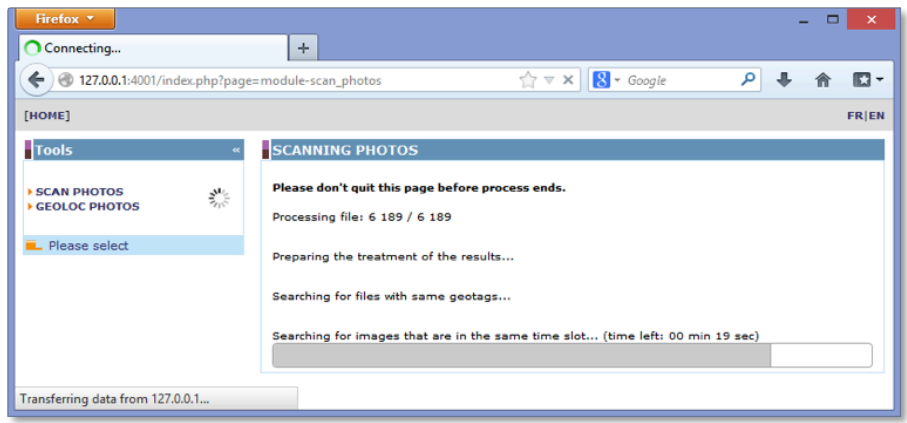

*Figure 5* – Scanning and Indexing Images

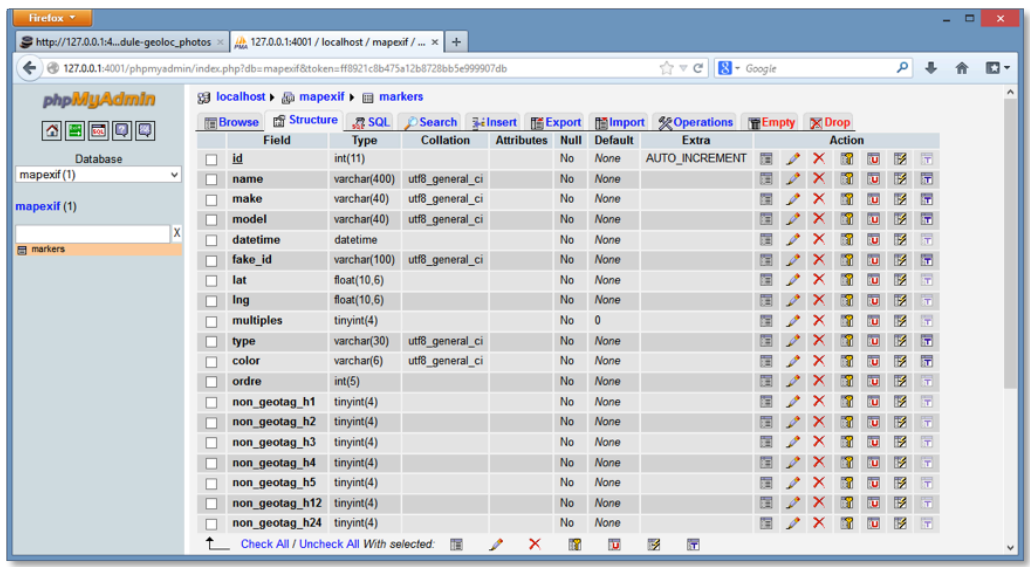

*Figure 6* – Structure of MapExif databases

As shown in Figure 7, after the scanning and indexing process, MapExif can show all images in its databases. In the example in Figure 8, we have 1148 images from more than 100 different devices.

**Filtering**

The objective of this functionality is to search for markers based on a zone filter. To do so, we first create a circle on the map thanks to the circle class of the google maps API (Google

Maps API). To find locations in the markers table that are within a certain radius of a given latitude/longitude, we use a SELECT statement based on the Haversine formularef). The Haversine formula (Appendix A) is used generally for computing great-circle distances between two pairs of coordinates on a sphere. The script can be found in Appendix A. Figure 9 shows a zone filter activated.

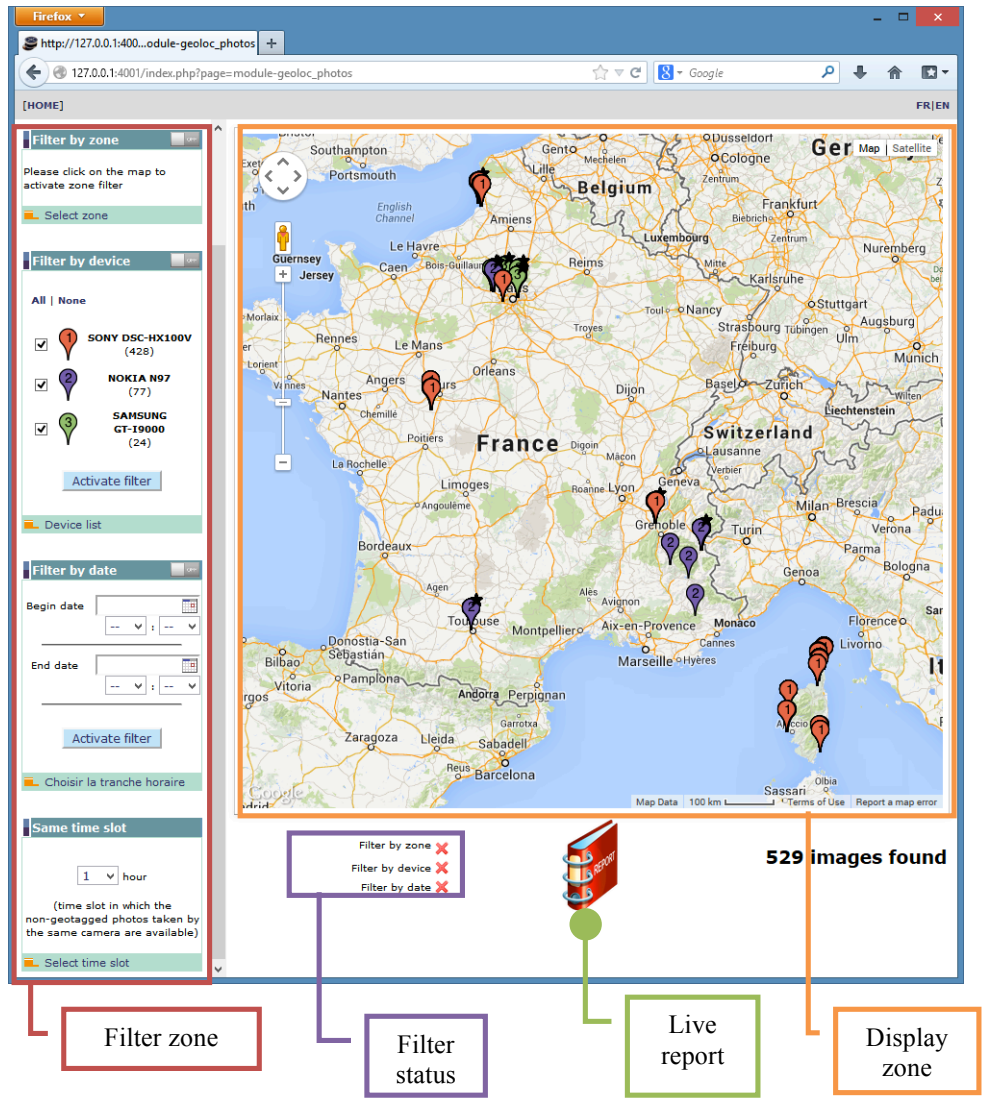

*Figure 7* - Description of the mapping interface

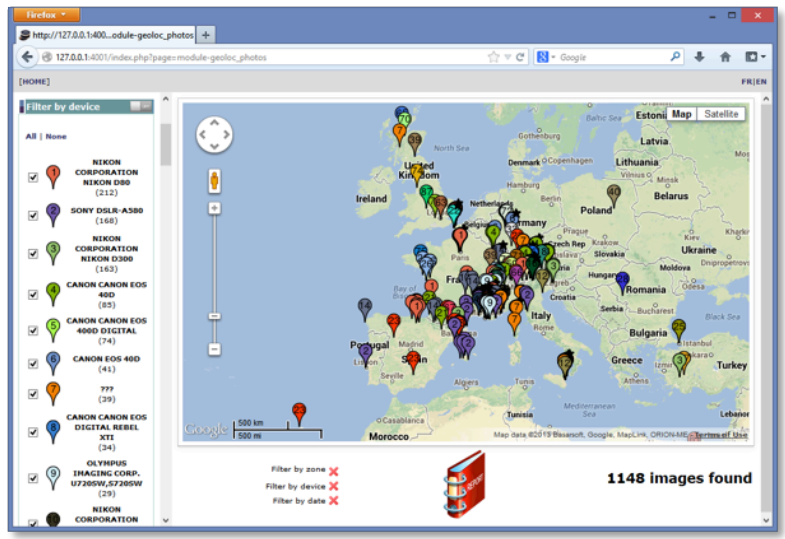

*Figure 8*- More than 100 different devices found

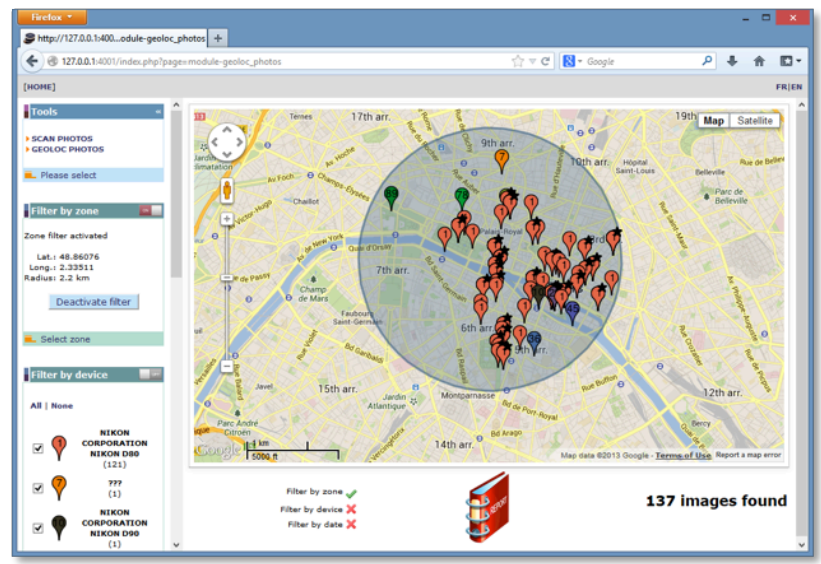

*Figure 9* - Zone filter activated

**Display option**

Based on the Google Maps API, satellite and street map views are also available in MapExif. The latter is of particular interest to investigators, allowing them to get an idea of places without having to go there. Figure 10 and 11 show these functionalities of MapExif.

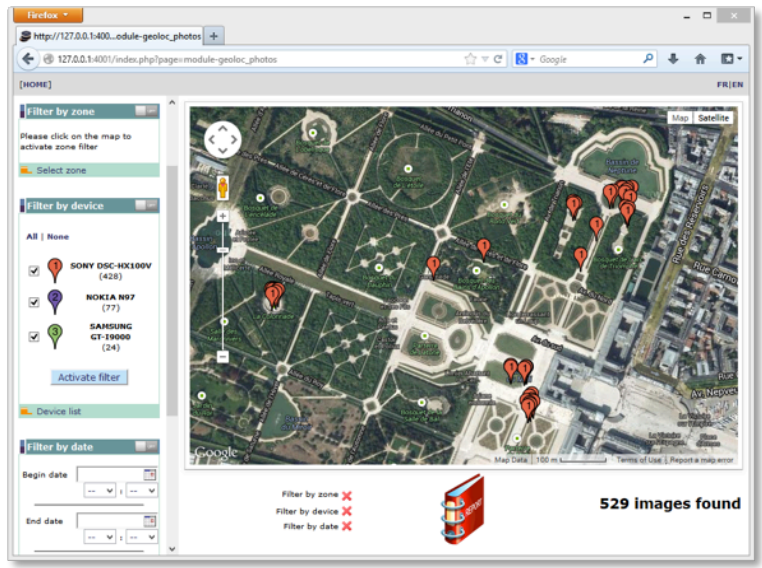

*Figure 10* - Satellite view

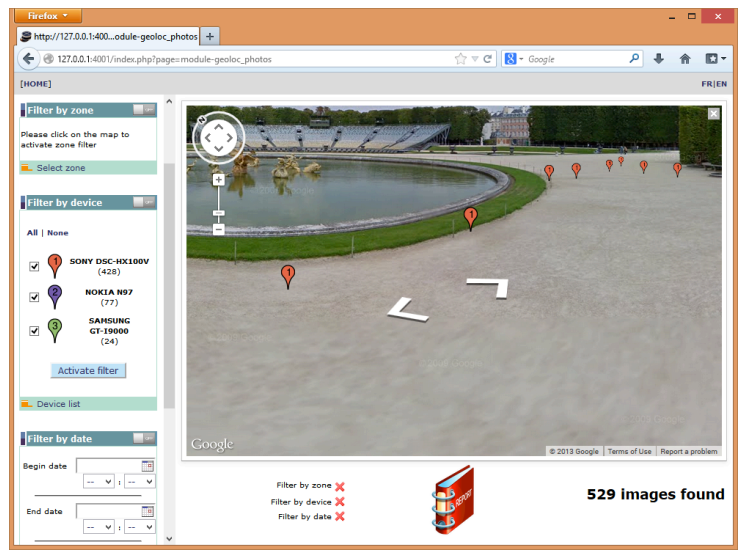

*Figure 11* - Street map view

**Picture spotting**

Regarding the markers, one of the most important operations is to allow investigators to quickly explore relevant images. This is very useful if there is a high density of markers in a small area. In this case, MapExif can show a thumbnail of the image and a short description is directly displayed when a marker is clicked (Figure 12). Another click on the thumbnail then allows the user to see the full size image with all its information (Figure 13).

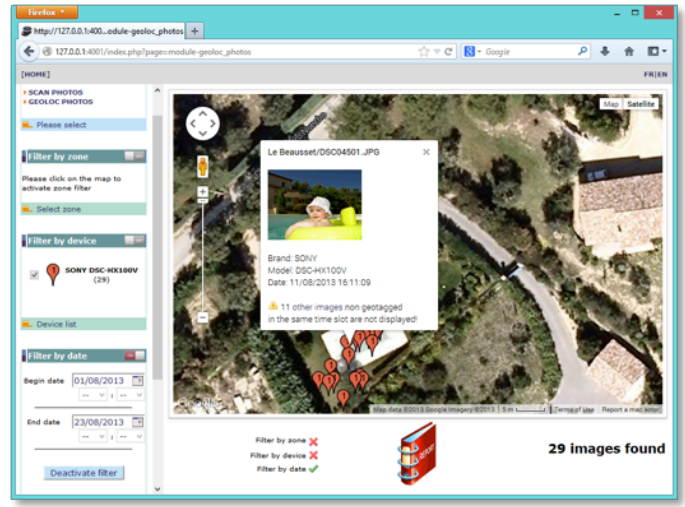

*Figure 12* - Miniature and brief information of the image

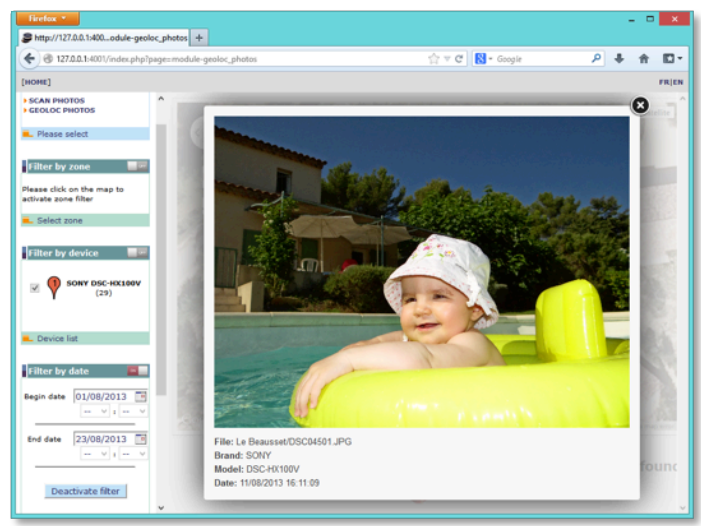

*Figure 13* - Image in full screen with all the information

In the event that there are two or more images having exactly the same GPS coordinates (e.g. in the case of a Wi-Fi-based positioning system), only one marker can be displayed, and the "reference image" is chosen arbitrarily. A star is added to the marker indicating that other images have the same geolocation and it is possible to view them by clicking on the warning information in the marker (Figure 14 and Figure 15).

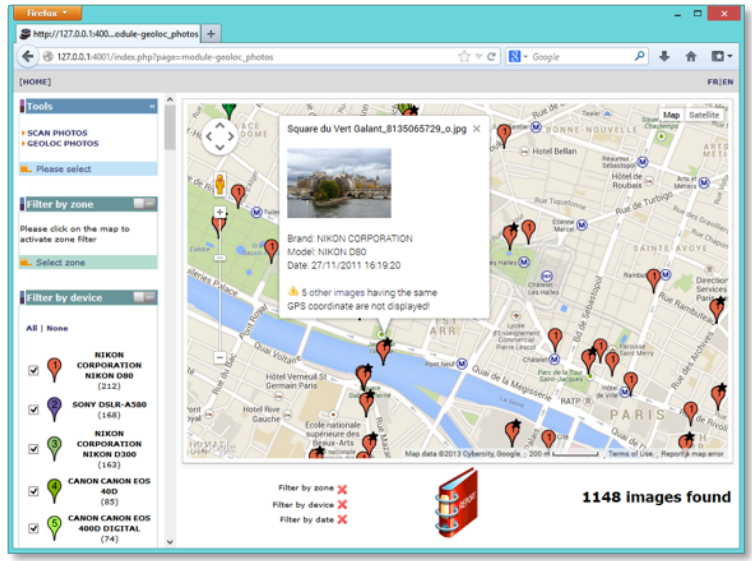

*Figure 14* - The selected marker indicates the location of 6 photos

only one image here      two or more images with the same coordinates

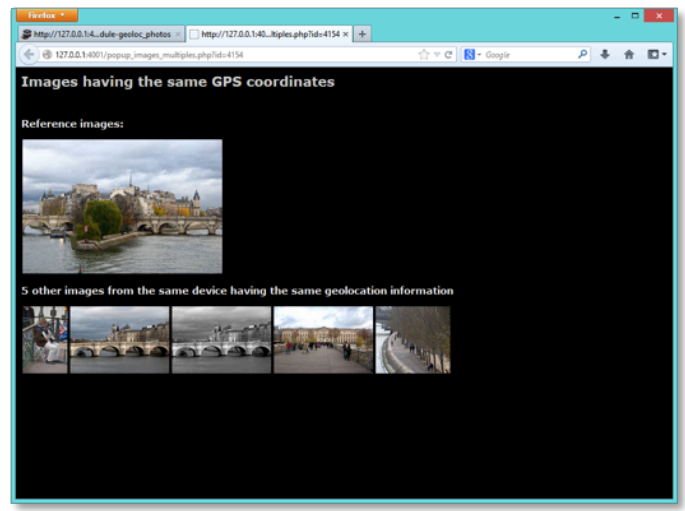

*Figure 15* - The reference image and the other five images with the same coordinates

**Non-geotagged images in the same time slot**

This is one of the most requested features by investigators because photos are not always geotagged and yet they still can be of paramount importance. Starting from a geotagged picture, MapExif shows other pictures taken by the same device in a chosen time slot, that are not geotagged. The connection made by MapExif does not guarantee that the proposed pictures have been taken in the same area as the reference geotagged picture. It is a mere probability that obviously decreases as the time slot gets larger. Examples on non-geotagged images can be found in Section 5.

**Live report**

During examination of the images, investigators can generate at any time a report that takes into account the current states of the filters. When changing these filters, the report is updated simply by refreshing the page. Using sample set of images, we can set the filter to all pictures taken by the device "NIKON D300". We can then generate the corresponding report (Figure 16). We can also narrow our report to a specific location (Figure 17).

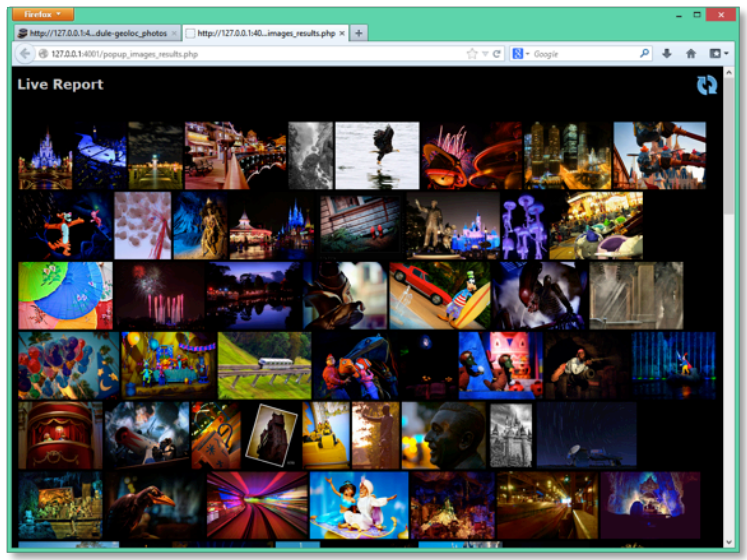

*Figure 16* – Live Report

**Experiments and Analysis**

**Case study**

In this section, we test our MapExif by exploring the following case study: A kidnapping has just occurred and as everyone knows the first hours are crucial. An amber alert is activated and a call for witnesses. For example, people who were around the scene at the time of the kidnapping may be asked to send the photos they have taken, because they can provide crucial information for investigators. It is possible to set MapExif to scan a mailbox and display on a map the photos that meet the right criteria (time, place ...).  a timeline report is also displayed. The investigator does not have to waste time opening and viewing multiple email attachments.

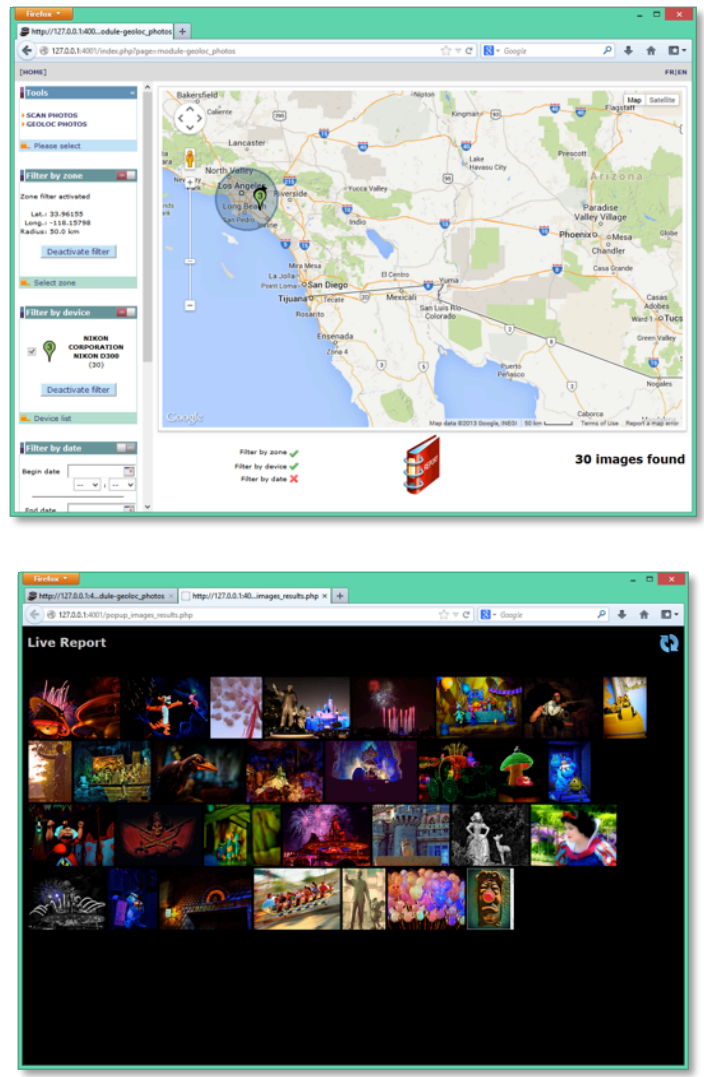

*Figure 17* –Live Report for Log Angeles area

If witnesses said that the criminals were driving a blue car, investigators could then  focus on photos found at the home of a suspect, and set zone and time filters to see if they could find a picture containing a blue car (Figure 18). If there is another image with no GPS coordinates taken by the same camera and in such a time interval defined, then MapExif would provide a warning (Figure 19). The "linked" image appears in the report and the investigators decide to examine it (Figure 20). After viewing this picture, investigators were convinced that this is the kidnappers' car (Figure 21).

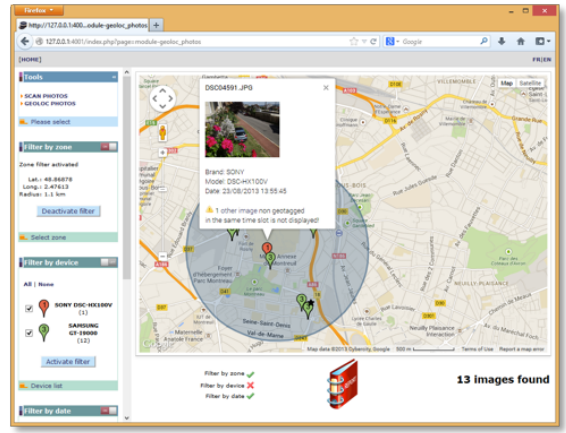

*Figure 18 –*

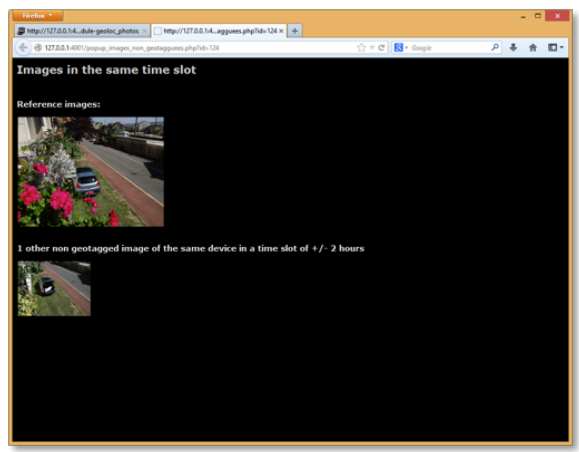

*Figure 19 –*

**Evaluation and Discussion of results**

    Geo-positional systems have gained in technological prominence in recent years. These systems provide location based information to applications which, in some cases, record geographic coordinates within Exif data stored in files. But can investigators rely on this information? From above, it is clear that various metadata from photos can provide useful information to forensic investigators.

The objectives for the development of MapExif are met. The tool is able to recognize the pictures without relying on extensions, to represent them on a map sorted by devices and to make the connection between a geotagged picture and non geotagged one within a given time slot. In addition, a live report is generated at any time, taking into account the defined filters (area, device, time).

However it is important to note that metadata should be used with caution when it comes to making decisions directly based on them. To illustrate this, as we saw earlier, the metadata describing the devices may not be very precise and two different devices of the same brand and model could be considered as one. As the metadata is often not protected, it is susceptible to manipulation. Metadata, for instance the name and details of an illegal image, can be changed maliciously and an innocent victim trapped by a malicious attacker.

Exif metadata (and therefore MapExif) can improve the quality of decision making in forensic investigations but they must be handled with care as their integrity can still not be guaranteed.

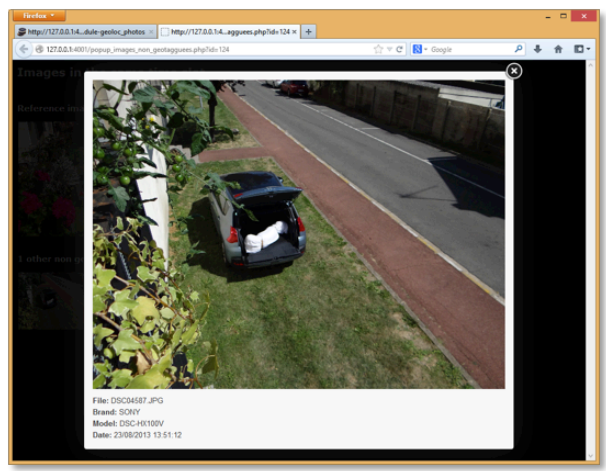

*Figure 20 –*

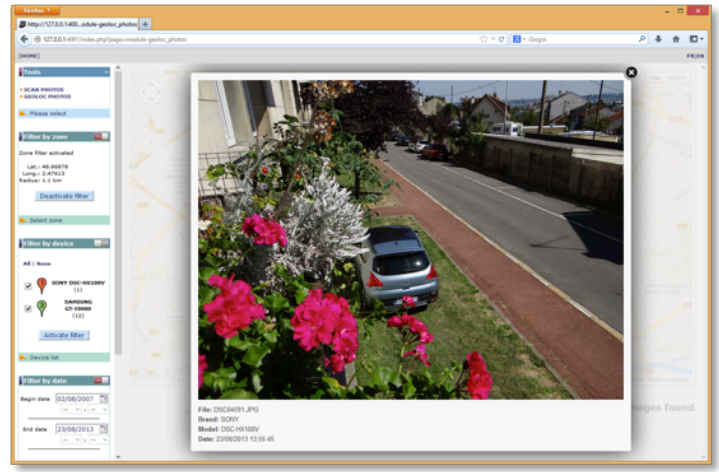

*Figure 21 –*

## Conclusions and future work

Right now MapExif is at its very beginning. It works on a single computer with local data. The next step will be a server-centralized server with crosscheck and links between different analyses of different investigators. For example a picture from a camera recovered during a search will have a serial number that matches with a stolen camera in another place. Thus the suspect could be charged with burglary.

The automatic scan of pictures on the web is also currently being tested. A crawler which has been set up with specific keywords is "patrolling" the web and sends to MapeExif every picture with metadata it finds. When a photo matches the selected filters, it is automatically sent to the investigator. It can be useful for example in case of intelligence gathering on terrorist activities, doing deep web searches on "specialized" websites.

We are working on the connections between isolated photos where no serial number is provided. An issue could be the legal implications - what kind of information can we store and work on. Is the picture considered to be personal information, or can it be just be linked to the device (and not to the person) as we consider that the brand, model of the device is not personal information. This opens up new perspectives for investigators.

# Appendix A            Programming Codes

## A.1. The following code shows how to extract GPS data and converting into decimal degree:

```php
//get the Hemisphere multiplier
$LatM  = 1;
$LongM = 1;
if ($exif["GPSLatitudeRef"] == 'S')
    $LatM = -1;
if ($exif["GPSLongitudeRef"] == 'W')
    $LongM = -1;

//get the GPS data
$gps['LatDegree']   = $exif['GPSLatitude'][0];
$gps['LatMinute']   = $exif['GPSLatitude'][1];
$gps['LatgSeconds'] = $exif['GPSLatitude'][2];
$gps['LongDegree']  = $exif['GPSLongitude'][0];
$gps['LongMinute']  = $exif['GPSLongitude'][1];
$gps['LongSeconds'] = $exif['GPSLongitude'][2];

//convert strings to numbers
foreach ($gps as $key => $value) {
    $pos = strpos($value, '/');
    if ($pos !== false) {
        $temp = explode('/', $value);
        $gps[$key] = $temp[0] / $temp[1];
    }
}

//calculate the decimal degree
$result['latitude']  = $LatM * ($gps['LatDegree'] + ($gps['LatMinute'] / 60) + ($gps['LatgSeconds'] / 3600));
$result['longitude'] = $LongM * ($gps['LongDegree'] + ($gps['LongMinute'] / 60) + ($gps['LongSeconds'] / 3600));
```

## A.2. Identify the different devices:

```php
if (isset($exif['Make']))
    $result['make'] = trim($exif['Make']); // brand of the device
if (isset($exif['Model']))
    $result['model'] = trim($exif['Model']); // model of the device
```

If some more information is available about the device, they are added to the fake id, so it is possible to distinguish between 2 devices as shown in the following code:

```php
if (isset($exif['SerialNumber'])) $result['optional_infos'] .= " | ".$exif['SerialNumber'];

if (isset($exif['OwnerName'])) $result['optional_infos'] .= " | ".$exif['OwnerName'];
if (isset($exif['LensInfo'])) $result['optional_infos'] .= " | ".$exif['LensInfo'];
if (isset($exif['LensMake'])) $result['optional_infos'] .= " | ".$exif['LensMake'];
if (isset($exif['LensModel'])) $result['optional_infos'] .= " | ".$exif['LensModel'];
if (isset($exif['LensSerialNumber'])) $result['optional_infos'] .= " | ".$exif['LensSerialNumber'];
```

## A.3. Generation of the Markers

```
/*
  GENERATION OF THE MARKERS, THE FOLLOWING CODE HAS BEEN SIMPLIFIED
*/

// call the xml file with will give the markers informations
downloadUrl("./ajax/module-geoloc_photos_genxml.php", function(data) {
var xml = data.responseXML;
var infos = xml.documentElement.getElementsByTagName("info");
var markers = xml.documentElement.getElementsByTagName("marker");
var marker_length = markers.length;
// creating markers browsing the xml file
for (var i = 0; i < marker_length; i++) {
  var name = markers[i].getAttribute("name");
  var id = markers[i].getAttribute("id");
  var brand = markers[i].getAttribute("brand");
  var model = markers[i].getAttribute("model");
  var fake_id = markers[i].getAttribute("fake_id");
  var date = markers[i].getAttribute("date");
  var point = new google.maps.LatLng(
     parseFloat(markers[i].getAttribute("lat")),
     parseFloat(markers[i].getAttribute("lng")));
  var marker = new google.maps.Marker({
   map: map,
   position: point,
   icon: icon_display,
   title: name + "\nBrand: " + brand + "\nModel: " + model + "\nDate: " + date + ""
  });
}
}

//AJAX FUNCTION
function downloadUrl(url, callback) {
  var request = window.ActiveXObject ?
  new ActiveXObject('Microsoft.XMLHTTP') :
  new XMLHttpRequest;
  request.onreadystatechange = function() {
    if (request.readyState == 4) {
     request.onreadystatechange = doNothing;
     callback(request, request.status);
    }
  };
  request.open('GET', url, true);
  request.send(null);
}
```

## A.4. Haversine

```
$query = "SELECT * ";
if ($filter_circle == 1)
    $query .= ", ( 6371 * acos( cos( radians($lat_input) ) * cos( radians( lat ) ) * cos(
radians( lng ) - radians($lng_input) ) + sin( radians($lat_input) ) * sin( radians( lat ) ) ) )
AS distance FROM markers HAVING distance < $radius_input";
```